\RequirePackage{fix-cm}
\documentclass[smallextended]{svjour3}       % onecolumn (second format)
\smartqed  % flush right qed marks, e.g. at end of proof
\usepackage{graphicx}
\usepackage{amsmath}
\usepackage{algorithm}
\usepackage{algpseudocode}
\usepackage{algorithmicx}
\usepackage{hyperref}
\usepackage{color}

\begin{document}

\title{Clustering articles based on semantic similarity}

%\subtitle{Do you have a subtitle?\\ If so, write it here}

\author{Shenghui Wang         \and
        Rob Koopman 
        }
%AS: Shenghui, Rob and myself spoke in Berlin that the short paper should be othered by you both. I have hardly contributed any substantial to it, and move my name to the acknowledgement if this is ok with you 
\institute{R. Koopman \at
              OCLC Research, Schipholweg 99, Leiden, The Netherlands \\
              Tel.: +31 71 524 6500\\
              \email{rob.koopman@oclc.org}           %  \\
%             \emph{Present address:} of F. Author  %  if needed
           \and
           S. Wang \at
              OCLC Research, Schipholweg 99, Leiden, The Netherlands \\
              Tel.: +31 71 524 6500\\
              \email{shenghui.wang@oclc.org}           %  \\
%             \emph{Present address:} of F. Author  %  if needed
			%\and 
            %A. Scharnhorst \at
             % DANS-KNAW, Anna van Saksenlaan 51, The Hague, The Netherlands \\                     %Tel.: +31 70 349 4450\\
           %   \email{andrea.scharnhorst@dans.knaw.nl}           %  \\
%             \emph{Present address:} of F. Author  %  if needed
}

\date{Received: date / Accepted: date}
% The correct dates will be entered by the editor

\maketitle

\begin{abstract}

Document clustering is generally the first step for topic identification. Since many clustering methods operate on the similarities between documents, it is important to build representations of these documents which keep their semantics as much as possible and are also suitable for efficient similarity calculation. 
%As part of the special issue ``Same data, different results?" this paper details the use of two standard clustering methods, the K-Means and the Louvain community detection algorithm to cluster articles in the so-called \textit{Astro} dataset. 
As we describe in~\cite{koopman2015_issi}, the metadata of articles in the \textit{Astro} dataset contribute to a semantic matrix, which uses a vector space to capture the semantics of entities derived from these articles and consequently supports the contextual exploration of these entities in \textit{LittleAriadne}. However, this semantic matrix does not allow to calculate similarities between articles directly. In this paper, we will describe in detail how we build a semantic representation for an article from the entities that are associated with it. % reverse engineer from the semantic matrix for the entities to a semantic matrix for the articles. %To be comparable to other approaches used by other groups, we construct from the semantic matrix various similarity matrices between articles. 
Base on such semantic representations of articles, we apply two standard clustering methods, K-Means and the Louvain community detection algorithm, which leads to our two clustering solutions labelled as OCLC-31 (standing for K-Means) and OCLC-Louvain (standing for Louvain). %, which we also displayed in \textit{LittleAriadne} \cite{koopman2015_ariadne}. 
In this paper, we will give the implementation details and a basic comparison with other clustering solutions that are reported in this special issue.

% \PACS{PACS code1 \and PACS code2 \and more}
% \subclass{MSC code1 \and MSC code2 \and more}
\end{abstract}

\section{Introduction}
\label{sec.intro}

Topics, sub-fields, specialities build the core in the self-organised process of scientific knowledge production~\cite{Bruckner1990}. There is a lot of ambiguity how to define these units of cognitive and social organisation~\cite{Sugimoto2015}, and an ongoing debate about how to extract them in an automatic, algorithmic way~\cite{intro}. Still, one way to identify \textit{topics} is to cluster documents. There are different ways to determine if two documents  address a related subject matter. Some well-known signals for a topical relatedness include citations (if one document cites another)~\cite{Garfield1963}, co-citations (if two documents are cited by a third document)~\cite{small1973}, bibliographic coupling (if two documents share a reference in their bibliography)~\cite{Glanzel1996}, and co-word linkages (if two documents share certain words)~\cite{Leydesdorff1989}. Each of these signals or traces can be used to construct a different matrix of relatedness or similarity between documents, based on which clusters of documents or topics can be identified. 

In the bibliometric literature advantages and disadvantages of different methods have been discussed in abundance. In general, one differentiates between citation-based and text-based metrics~\cite{boyack2013}. Although words are expected to be less codified than cited references, we share the belief that words, especially those in titles and abstracts, do carry a certain amount of a knowledge claim made by a paper~\cite{leydesdorff2006measuring}. Hence, in accordance to the programme of cognitive scientometrics~\cite{ripcourtial1984} and more recent full-text based bibliometric studies~\cite{boyack2013}, we state that if two documents share enough lexical information, they are considered to be related. 

For the clustering approaches detailed in this paper, we rely on a new semantic representation of articles to determine their similarities. Both the underlying method and an interactive search interface based on it has been named \textit{Ariadne}~\cite{koopman2015_chi,koopman2015_issi}. Our approach has great resemblance to methods used in information retrieval, in as such that it operates in a word space. But in difference to methods based on Salton's word space of documents, we use information from all elements of a document (in our case, an article), and create a word space for all those elements or entities. The motivation for this is based on the assumption that using information from many different elements of an article provides a more accurate semantic representation of this article. We consequently assume that this also improves the basis on which the similarity/relatedness between articles is determined. When we use entities such as authors, journals, subjects-headings, or references we simultaneously search for semantic similarity/relatedness along perspectives of a social (authors), communicative (journals as publication venue), or knowledge exchange (references) organisation of scientific knowledge production. 

Our \textbf{research questions} are therefore (a) whether we could reconstruct a valid semantic representation for articles from all the entities they are associated with and (b) identify article clusters using standard methods based on such a semantic representation.   

In this paper, we first describe how to represent the semantics of articles based on the entities that are involved with these articles. Then we briefly introduce two standard clustering methods, K-Means and Louvain community detection algorithm before reporting the implementation details. At the end, we compare our two solutions with the other clustering solutions reported in this special issue and conclude the paper.  

%We give a summary of the \textit{Ariadne} approach in the next section. For more a detailed description we would like to refer to other papers~\cite{koopman2015,koopman2015c}.

%AS: the following is very nice! can we use this for the long method article for IJDL
\section{From semantics of entities to semantics of articles}
\label{sec.integration}
For our approach, we adopt the notion of \textit{Statistical Semantics}~\cite{furnas1983,weaver1955} based on the assumption of ``a word is characterized by the company it keeps''~\cite{firth1957} or in Linguistics the \textit{Distributional Hypothesis}~\cite{harris1954,sahlgren2008}: words that occur in similar contexts tend to have similar meanings. In \textit{Ariadne}, we extend words to entities (such as authors, journals, subjects, citations) so that each entity is indexed by a vector in a semantic space reflecting their lexical context, i.e., their co-occurrences with certain terms (including topical terms extracted from title and abstract plus user-defined subjects)~\cite{koopman2015_ariadne}. 

The resulting entity-term co-occurrence matrix could become extremely big and sparse which makes any computation on top of it very expensive and impractical. Thanks to Random Projection~\cite{Achlioptas2003671,johnson84extensionslipschitz}, we can dramatically reduce the dimensionality of this semantic space, obtaining  a much smaller and manageable sized \textit{semantic matrix} yet keeping the semantics of the entities as much as possible. With all entities represented as vectors in the same semantic space, it is possible to compute the distance or relatedness between any pairs of entities, no matter which types they are. Such freedom is a unique feature of \textit{Ariadne}. It provides a contextual view about an entity or a query as a start of an exploratory journey.  For  a more detailed description please refer to other papers~\cite{koopman2015_chi,koopman2015_ariadne}.

In the semantic matrix each article contributes to the semantics of individual entities. When executed over a big corpus the statistics are reliable to calculate the similarity between entities, However, from this semantic matrix, we cannot directly calculate similarities between articles. 
%@Shenghui: I don't really get this: I thought, it is not just a matter of statistics, but also that the articles are kind of contribute in a weighted different way, and so it is not possible to add them up. "Unfortunately, the articles themselves could not be represented in the same way that individual entities are represented due to the lack of statistics."  

To be able to cluster articles, and thus be comparable to the other methods, we first construct an integrated representation of an article from the entities associated with it. To do so, for each article, we look up all entities associated with this article in the Semantic Matrix. Consequently we obtain a set of vectors $V=\{\vec v_1, \ldots, \vec v_{n}\}$ for each article, where $n$ is the number of entities associated with it and $\vec v_i$ is the vector for entity $e_i$. These entities can be the authors, subjects, journal, citations, topical terms (extracted from its title and abstract), etc. Each article is represented by a unique set of vectors. The size of the set $n$ can vary, but each of the vectors inside of a set has the same length, in our case 600 (see~\cite{koopman2015_ariadne} for more details). 

For each article we now build a new vector $\vec v'$, the weighted centroid of its constituent vectors: %. Its components $v^{'}_i$ consist of a weighted average of these entity vectors, i.e.,
%@Shenghui: I got problems with the Tex commands here, please put this right. We have vectors (with an index i) and we have vector components (with an index j). The equation describes how one specific component in a vector for an article needs to be calculated. I will try to get this handwritten on paper tomorrow. Notation is not yet 100% I feel.
\begin{equation} 
\label{eq.integration}
\vec v^{'}= \frac{\sum_{i=1}^{n} w_i \cdot \vec v_i}{\sum_{i=1}^{n}w_i},
\end{equation}
where $w_i=log(N/f_i)^3$, $N$ is the total number of articles and $f_i$ is the number of articles which contain the entity $e_i$. With this specific weighting frequent entities are heavily penalized to have little contribution to the resulting representation of the article. In the end, each article is represented by a vector of 600 dimensions. % which becomes a row in a new matrix $M$ with the size of $111,616 \times 600$.

\paragraph{Feature selection} We extend our results published in~\cite{koopman2015_issi} by putting the citations as additional entities in the Semantic Matrix (see~\cite{koopman2015_ariadne} for more details). In order to see which role the citation information plays in terms of clustering, we will experiment by including or excluding citation vectors when computing the semantic vectors for articles (Equation~\ref{eq.integration}). So, for each article, we generate 3 vectors, one is a weighted average of everything but citations (i.e., topical terms, subjects, authors, and journals, the same in~\cite{koopman2015_issi}), one is a weighted average of only citation entities, and one is a weighted average of all types of entities. In Section~\ref{sec.kmeans}, we will report the comparison results.

%K-mean is a standard clustering techniques  (ref?)  for large vector spaces, which is often applied in data-mining problems. For this technique one can rely on standard algorithmic implementations. The second algorithm we apply is the so-called Louvain method, which uses modularity features of the network~\cite{louvain}. Since its introduction the method has become very popular, also in scientometrics~\cite{Zhang2010}. The algorithm is also used by xxx in this special session, and this was another reason to chose it.   
%@Shenghui: please add which ones use Louvain, give citation place holders

\section{Standard cluster algorithms}

Once the article vectors are generated, the next step is to identify clusters of articles. Various clustering methods can be applied. We mainly experiment with K-Means because it is a simple and highly scalable clustering method which directly operate on the vectorial representations of the articles. Our goal is to check whether such semantic representations yields sensible clusters. 

Network-based clustering methods are well used in the scientometrics community. Therefore, we also try to solve the clustering problem from a network point of view. As a further process of such semantic representations, we transform the similarities calculated based on such vectorial representations to a similarity network of articles from which communities (clusters) could be detected. We choose to apply the Louvain community detection method~\cite{louvain} as it is widely used in the scientometrics community but mostly applied to citation-based data models. We are interested to check whether the Louvain method could also find communities based on semantic similarities of the articles, instead of citations between them. 

%We Instead of  directly operating on the vectorial representations of articles, we also appl the so-called Louvain community detection method~\cite{louvain}, which operates on an article network derived from the vectorial representations of articles. 
We now briefly describe these two standard algorithms and the implementation details on our dataset.

%. But we also created a clustering solution using the Louvain method. The reasons were the popularity of this method in the scientometrics community and its use by other teams in this special issue~\cite{ecoom2015}. 

\subsection{Clustering using K-Means}
\label{sec.kmeans}

The K-Means algorithm is one of the simplest unsupervised learning algorithms that solves the well defined clustering problem~\cite{mackey2003,datamining}. It scales well to large number of samples and has been used across a large range of application areas in many different fields including scientometrics~\cite{boyack2005mapping}. 
%@Shenghui: can we find a scientometric paper using it? I also recall that Kevin mentioned that although it is known in the scientometrics community, there have been reasons why it is not that popular. I will ask him for pointers. Then we could write something as: Although in scientometrics other clustering techniques are more popular and k-means often is discarded because of ... it still is a standard algorithm, used for clustering document spaces but than by researchers outside of scientometrics ... ... not yet very elegant (ASJan 10)

Given a set of data points or observations ($x_1, x_2, \ldots, x_n$), where each data point is characterized by a d-dimensional real vector, k-means clustering aims to partition the $n$ data points into $k(\leq n)$ sets or clusters $S = \{S_1, S_2, \ldots, S_k\}$ so that the Within-Cluster Sum of Squares (WCSS) is minimized. In other words, the objective of the K-Means algorithm is to find
%@Sheghui: equation (1) is what? The function WCSS? which is to be optimized? please make an equation out of it. I know the formula comes from wikipedia - but it is not yet clear (to me) 
%@Andrea: the part from the first sigma is WCSS, argmin_S is the S which gives the minimum WCSS. It is to optimise the WCSS and find the S. 
\begin{equation}
\underset{\mathbf{S}} {\operatorname{arg\,min}}  
\sum_{i=1}^{k} \sum_{\mathbf x \in S_i} \left\| \mathbf x - \boldsymbol\mu_i \right\|^2 
\end{equation}
where $\mu_i$ is the centroid (mean) of points in $S_i$.

This algorithm requires the number of clusters to be specified a priori. It starts with an initial set of $k$ centroids $m_1^{(1)},\ldots,m_k^{(1)}$ and proceeds by alternating between two steps~\cite{mackey2003}:
\begin{description}
\item{\textbf{Assignment step}}: Assign each data point to the cluster whose mean yields the least WCCS.\footnote{Since the sum of squares is the squared Euclidean distance, this is intuitively the ``nearest" mean.}
\begin{equation}
S_i^{(t)} = \big \{ x_p : \big \| x_p - m^{(t)}_i \big \|^2 \le \big \| x_p - m^{(t)}_j \big \|^2 \ \forall j, 1 \le j \le k \big\},
\end{equation}
where each $x_p$ is assigned to exactly one $S^{(t)}$, even if it could be assigned to two or more of them.
\item{\textbf{Update step}}: Calculate the new means to be the centroids of the data points in the new clusters.
\begin{equation}
m^{(t+1)}_i = \frac{1}{|S^{(t)}_i|} \sum_{x_j \in S^{(t)}_i} x_j 
\end{equation}
\end{description}
The algorithm converges when the assignments no longer change, which leads to a (local) optimum while the global optimum is not guaranteed. 

The Mini Batch K-Means~\cite{minibatch} is a variant of the K-Means algorithm which uses mini-batches to reduce the computation time, while still attempting to optimize the same objective function. The algorithm takes small batches (randomly chosen) of the dataset for each iteration. It then assigns a cluster to each data point in the batch, depending on the previous locations of the cluster centroids. It updates the locations of cluster centroids based on the new points from the batch. The update is a gradient descent update, which is significantly faster than a normal Batch K-Means update. 
%Formally, the algorithm is presented in Algorithm~\ref{alg.minibatch}.
% \begin{algorithm}
% \caption{Mini Batch K-Means~\cite{minibatch}}
% \label{alg.minibatch}
% 1: Given: $k$, mini-batch size $b$, iterations $t$, data set $X$\\
% 2: Initialize each $c \in C$ with an $x$ picked randomly from $X$\\
% 3: $v \gets 0$\\
% 4: for $i = 1$ to $t$ do\\
% 5: \hspace{5mm}$M \gets b$ examples picked randomly from $X$\\
% 6: \hspace{5mm}for $x \in M$ do\\
% 7: \hspace{10mm}$d[x] \gets f(C, x)$ // Cache the center nearest to $x$\\
% 8: \hspace{5mm}end for\\
% 9: \hspace{5mm}for $x \in M$ do\\
% 10: \hspace{10mm}$c \gets d[x]$ // Get cached center for this $x$\\
% 11: \hspace{10mm}$v[c] \gets v[c] + 1$ // Update per-center counts\\
% 12: \hspace{10mm}$\eta \gets \frac{1}{v[c]}$ // Get per-center learning rate\\
% 13: \hspace{10mm}$c \gets (1 - \eta )c + \eta x$ // Take gradient step\\
% 14: \hspace{5mm}end for\\
% 15: end for
% \end{algorithm}

Using mini-batches drastically reduce the amount of computation required to converge to a local solution, but the quality of the results is reduced. In practice this difference in quality can be quite small~\cite{Be2013}. Therefore, we choose to use a Mini-Batch K-Means implementation provided by an open source machine learning library to cluster the articles in the \textit{Astro} Dataset, where  each article is a data point in the 600 dimensional semantic space, as described in Section~\ref{sec.integration}.

\subsection{Clustering using the Louvain method for community detection}
\label{sec.louvain}

We consider each article as a node in a network, and there is a link between two articles when they are highly similar. Practically in our case, we connect each article to its top 40 the most similar/related articles based on the cosine similarities calculated from their vectorial representation. This results in an article similarity network where clusters or communities could be detected. The task is to partition the network into communities of densely connected nodes, with no or sparse connections between the nodes belonging to different communities. 

The Louvain method~\cite{louvain} is a simple, efficient and well-accepted method for identifying communities in large networks. It is widely used in many applications in different domains including scientometrics~\cite{Zhang2010185,ecoom2015,Zhang2010}. We apply it to see how well it performs on a similarity network rather than a citation-based network as what the ECOOM team reported in this special issue. 

The method itself is a greedy optimization method that attempts to optimize the ``modularity" of a partition of the network. Modularity is a scale value between -1 and 1 that measures the density of edges inside communities compared to edges outside communities. It is defined as~\cite{modularity2}: 
% \begin{equation}
% Q = \Sigma_{c_i \in C} [\frac{|E_{c_i}^{in}|}{|E|} - (\frac{2|E_{c_i}^{in}| + |E_{c_i}^{out}|}{2|E|})^2]
% \end{equation}
% where $C=\{c_1, \ldots, c_k\}$ is the set of all the communities, $|E_{c_i}^{in}|$ is the number of edges between nodes in community $c_i$, $|E_{c_i}^{out}|$ is the number of edges from the nodes in community $c_i$ to the nodes outside of $c_i$, and $|E|$ is the total number of edges in the network.
\begin{equation}
Q= \frac{1}{2|E|}\Sigma_{ij}\bigg[A_{ij} - \frac{k_i k_j}{2|E|}\bigg] \delta (c_i,c_j)
\end{equation} 
where $|E|$ is the total number of edges in the network, $k_i$ is the degree of node $i$, $A_{ij}$ is an element of the adjacency matrix (e.g., the weight of the edge between $i$ and $j$), and $c_i$ is the community to which node $i$ is assigned, and the $\delta$ function is 1 if $c_i=c_j$ and 0 otherwise. 

The optimization is performed in two steps iteratively. In the first phase, the method looks for ``small" communities by optimizing modularity locally. Each node is initially assigned to a different community, i.e., there are as many communities as there are nodes. Then, for each node $i$ the gain of modularity is calculated by moving $i$ from its own community into the community of each neighbour $j$ of $i$. After this value is calculated for all communities $i$ is connected to, $i$ is placed into the community that resulted in the greatest modularity increase. If no positive gain is possible, $i$ remains in its original community. This process is applied repeatedly and sequentially to all nodes until no modularity improvement can occur and then the first phase is complete.

In the second phase, it aggregates nodes belonging to the same community and builds a new network whose nodes are the communities from the previous phase. Then the first phase can be re-applied to this new network. This way, it iteratively optimizes local communities until a maximum of global modularity is reached. 

Compared to K-Means, the advantage of using the Louvain method is that the number of partitions or clusters is decided by the data itself. Similar to K-Means, the Louvain method is also an approximate method which does not really guarantee a global maximum of modularity. But it is highly scalable and often produces good approximation of the optimal communities. 

\section{Experiments}

We applied the above-mentioned two clustering methods to the \textit{Astro} dataset, which contains 111,616 articles in astronomy and astrophysics from 2003 to 2010 (please see~\cite{intro} for a full description of the dataset).

\subsection{Experiments with K-Means}

\subsubsection{Determining $K$ based on a pseudo-ground-truth}
%it was originally used for determining K, but it became a quality measure for other results too.
%AS Jan 11: to be able to judge your own solution, you take ensembels of other solutions as reference system

Evaluating clustering results or detected communities is a complex problem. The results could be presented to experts who decide whether each cluster or community is valid or not. Alternatively, a ground truth, i.e., a reference cluster or community allocation, could be used to measure how well the clustering solution fits the ground truth. Unfortunately, either way is extremely labour intensive if not  impossible in our case.  

%A more practical problem while applying K-Means is the choice of $k$. Prior knowledge of the data is useful to find an optimal choice of $k$ which should strike a balance between maximum compression of the data using a single cluster, and maximum accuracy by assigning each data point to its own cluster. Unfortunately, the prior knowledge or ``ground truth" is not available in general, but 

This causes a practical problem while applying K-Means. A ground truth, or prior knowledge of the data, would help to determine one of the most important parameters for K-Means, the choice of $k$. The lack of ground truth forces us to determine $k$ pragmatically.

\begin{figure}[t]
%\label{fig.silhouette}
\centering
\includegraphics[width=.7\linewidth]{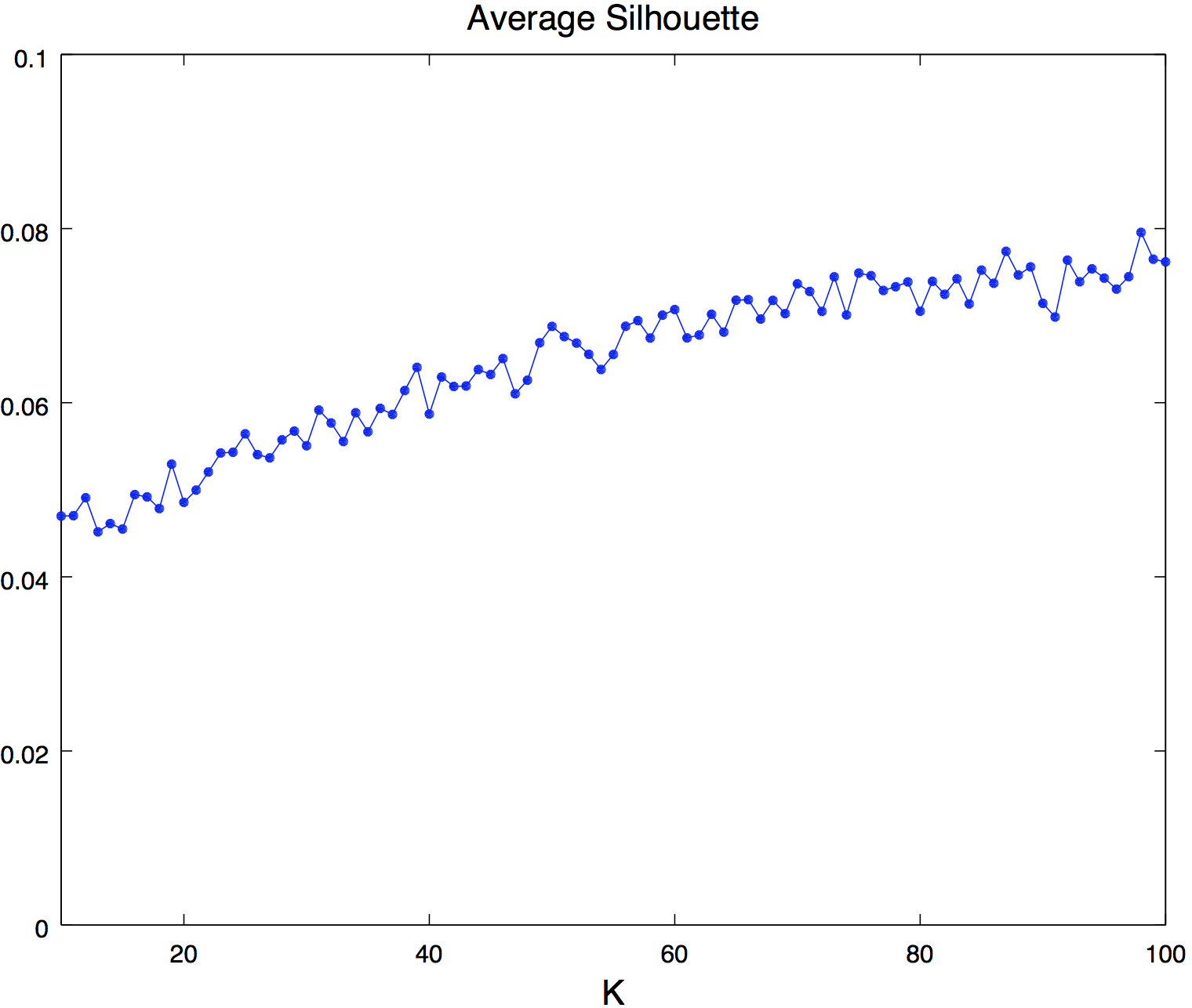}
\caption{Average Silhouette over 20,000 random chosen samples, with $k$  from 10 to 100 \label{fig.silhouette}}
\end{figure}

The average silhouette of the data~\cite{Rousseeuw1987} is a measure which could be used for determining $k$. The silhouette measures how closely a data point is matched to other data points within its cluster and how loosely it is matched to data points of the neighbouring cluster, i.e. the cluster whose average distance from the data point is lowest. The silhouette ranges from -1 indicating a wrong assignment to 1, an appropriate one while scores around zero indicating overlapping clusters. We calculated the average silhouette of a sample of 20,000 data points with $k$ from 10 to 100. As shown in Figure~\ref{fig.silhouette}, although slowly climbing the average silhouette scores are still around zero. This means that any numbers of clusters from this dataset are highly overlapping and a clear boundary between clusters seems not possible. This may reflect the intrinsically intertwined scientific communications between different topics. Another possible reason is that these articles may focus on different topics of the astrophysical domain, but they might still use the overlapping vocabulary which makes a clear distinction based on lexical information difficult to detect. 
% * <shenghui.wang@gmail.com> 2015-12-15T15:02:35.824Z:
%
% > This means that any numbers of clusters from this dataset are highly overlapping and a clear boundary between clusters seems not possible.
%
% come back to this in the discussion/conclusion
%
% ^.

Since there are already a couple of clustering solutions on the same dataset from different research teams, we could build a \textit{pseudo-ground-truth} based on the consensus of the available clustering solutions. We collected four clustering solutions, namely CWTS-C5, UMSI0, ECOOM-BC13 and STS-RG. Across all these four solutions, there are 93,986,261 pairs of articles, involving 96,072 articles (86\% of the whole dataset), are always in the same clusters. We use these shared pairs as the pseudo-ground-truth. It is by no means the real ground truth, but a consensus we could use to tune our $k$ to make a best guess. 

As Table~\ref{tab.solutions} shown, CWTS-C5 clusters provide the least number of article pairs while has the biggest proportion which is shared with the other three solutions. While the STS-RG clusters are quite the opposite: producing more than 940K article pairs but only 10\% of which are shared with others. It is mainly due to its largest 3 clusters which already contain 61\% of the whole data set. They produce a large amount of within-cluster article pairs. But because these articles are in the same clusters, the total amount of shared pairs is not reduced much by including the STS-RG clusters. Note that the STS-RG clusters are generated using a rather different method from those used by the other three~\cite{velden2015comparison}. Without the STS-RG clusters, there are 140M shared pairs and 100K articles involved. However, as we find that including them does not have much effect on the choice of $k$, we decided to include the STS-RG clusters to build our pseudo-grounth-truth. %, as shown in Figure~\ref{fig.measure}. %and 193 clusters only contain one articles

This simple comparison presented in Table~\ref{tab.solutions} also suggests there might be a core set of articles whose cluster assignments are rather stable no matter which clustering method is used. Therefore, we argue that this set of 93 million \textit{shared pairs} involving 96K articles could be used to evaluate new clustering solutions, such as our own Louvain results. 

%100167 articles, 140,155,810 pairs

\begin{table}
\caption{Statistics of the four clustering solutions for the pseudo-ground-truth}
\label{tab.solutions}       
\begin{tabular}{lllc}\\\hline
 & \#cluster & \#total pairs & of which are shared with the others \\\hline
CWTS-C5 & 22  & 337,151,232 & 28\%\\
UMSI0 & 22  & 453,492,311 & 21\%\\
ECOOM-BC13 & 13 & 498,846,580 & 18\%\\
STS-RG & 556  & 940,553,592& 10\% \\\hline
\end{tabular}
\end{table}

With this pseudo-ground-truth, we are looking for an optimal $k$. On one hand these $k$ clusters agree the most with the other four solutions, i.e., reproducing the most shared pairs. On the other hand large clusters are penalised if they put irrelevant articles into the same clusters. Formally we measure the precision ($p$) and recall ($r$) as follows:
\begin{equation}
p=\frac{\#article\ pairs\ in\ common}{\#total\ produced\ pairs}, \\
r=\frac{\#article\ pairs\ in\ common}{\#total\ shared\ pairs}
\end{equation} 
where ${\#total\ shared\ pairs}$ is the total number of the article pairs in the pseudo-ground-truth, i.e. 93 million, $\#total\ produced\ pairs$ is the total number of within-cluster article pairs produced by the $k$ clusters, and $\#article\ pairs\ in\ common$ is the number of article pairs which are produced by the $k$ clusters and also shared by the other four solutions. A high $p$ means a large proportion of produced article pairs are agreed by the other four solutions, while a high $r$ indicates that a large proportion of the shared pairs in the pseudo-ground-truth are produced by the $k$ clusters. A recall of 100\% can be reached by putting all articles in one cluster, but that would give a very low precision, as majority of the within-cluster article pairs are not agreed by the other four solutions. Many small clusters could improve the precision as they only contain the articles which are considered to be in the same cluster by the other four solutions, however, many potentially related articles are distributed in different clusters which damages the recall. 

To balance between $p$ and $r$, we calculate the $F1$ measure\footnote{\url{https://en.wikipedia.org/wiki/F1_score}} as widely used in the Information Retrieval community: 
\begin{equation}
F1= 2 \times \frac{p \times r}{p + r}
\end{equation}
Furthermore, under the similar situation with respect to $F1$, we are aiming at a reasonably higher level of abstraction, i.e., the larger clusters the better, provided that a reasonable number of irrelevant articles are included. Therefore we reward bigger cluster by adding a parameter of the average size of the clusters into the calculation. Therefore our final score for a set of clusters is calculated as:  
\begin{equation}
adjustedF1=F \times \log(avgSize)
\end{equation}
We therefore choose the best $k$ which gives the highest $adjustedF1$ score. As mentioned before, we will later use the $adjustedF1$ score to evaluate the clustering results from the Louvain method as well. 

\subsubsection{K-Means clustering results}

As mentioned in Section~\ref{sec.integration}, we build for each article  three vectorial representations: one averaging the semantic vectors of all entities, one with all entities except citations and one with only citation entities. We now search for the best $k$ for these three representations of articles.

\begin{figure}
\centering
\includegraphics[width=.8\linewidth]{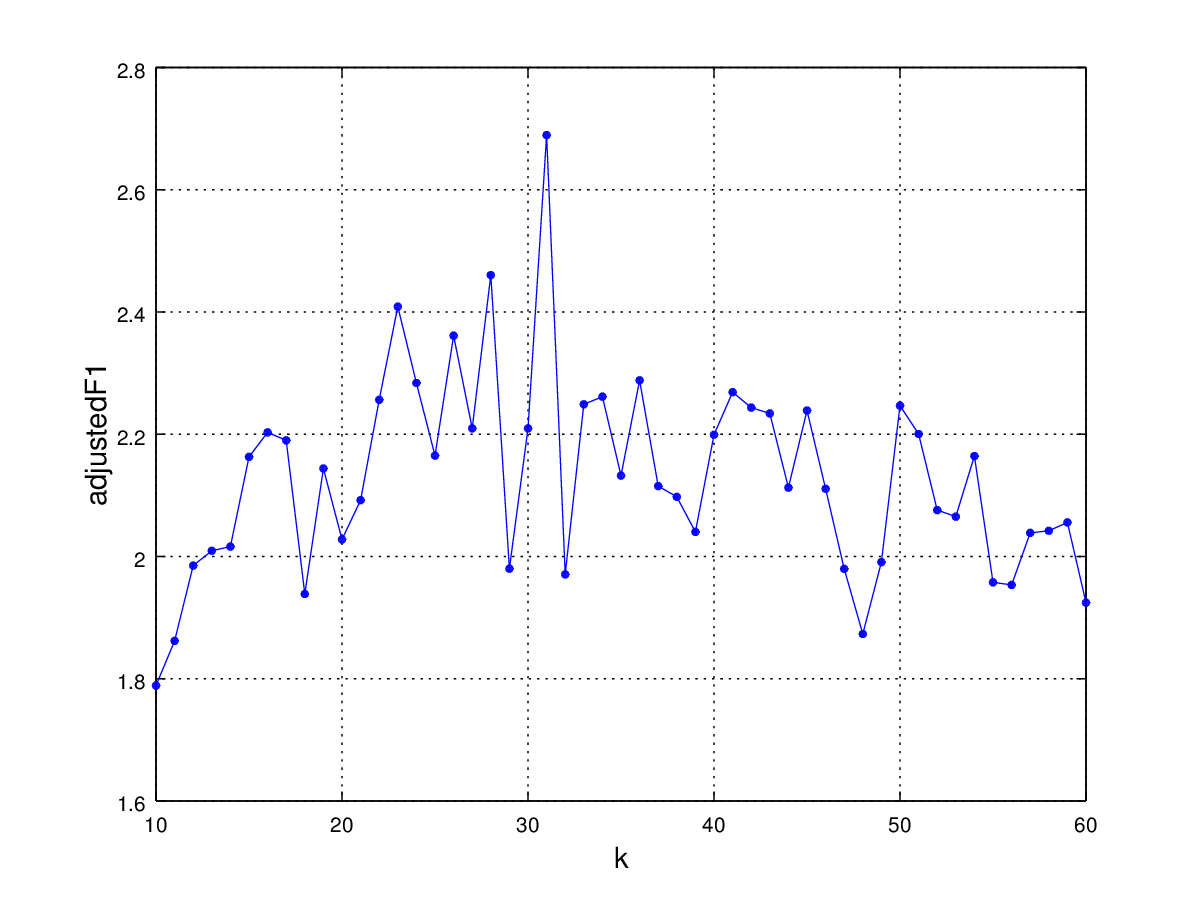}
\\
(a) all entities\\
\includegraphics[width=.8\linewidth]{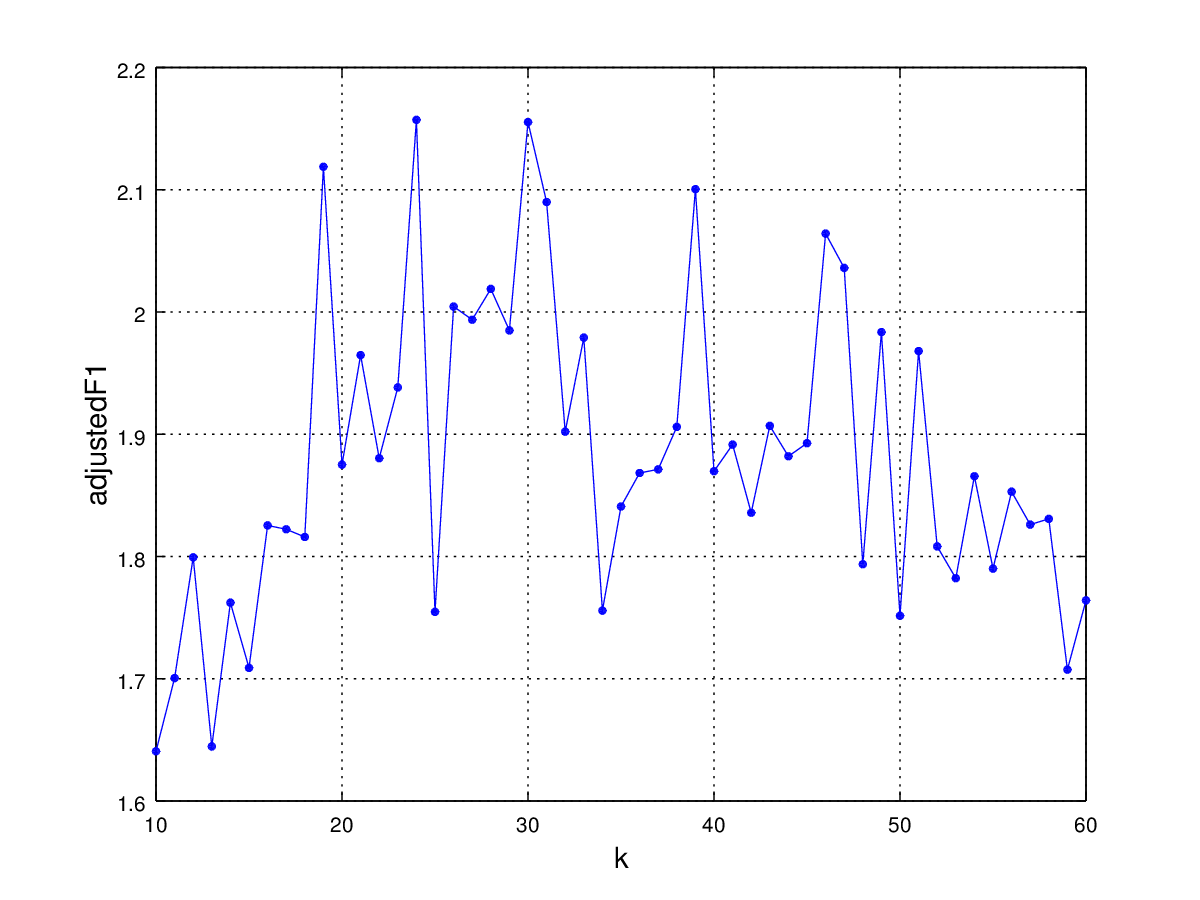}
\\
(b) no citations \\
\includegraphics[width=.8\linewidth]{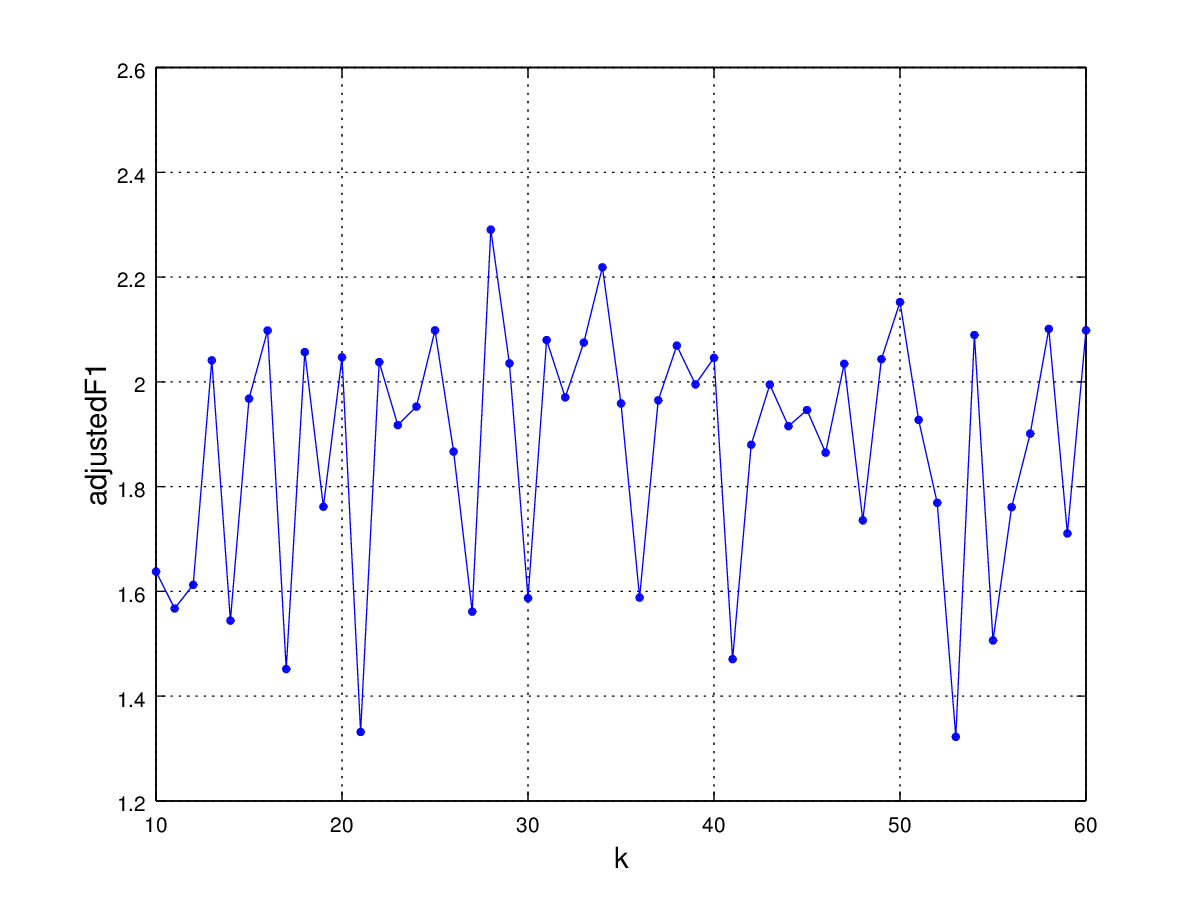}
\\
(c) only citations
\caption{Looking for the best $k$ based on $adjustedF1$, using different sets of entities \label{fig.measure}}
\end{figure}

The K-Means algorithm is sensitive to the initialization step, i.e. where the $k$ centroids are initially positioned. Therefore, for $k$ from 10 to 60, we ran 10 times the Mini Batch K-Means algorithm provided in the scikit-learn python library\footnote{\url{http://scikit-learn.org/}}  and chose the best solution which has the minimum WCSS. Then we used the $adjustedF1$ measure to evaluate our solutions against the pseudo-ground-truth. The $adjustedF1$ scores are plotted against $k$ in Figure~\ref{fig.measure}. 

If using all entities, the score climbs up until $k$ is around 30 then decreases, with $k=31$ giving the highest score. Therefore, we chose $k=31$ as the best $k$ if all entities are used for article semantic representation. Similarly, we found the best $k=28$ if only citations are used and $k=24$ if no citations are used. However,  Figure~\ref{fig.measure} (b) presents, if using no citations, there are much bigger fluctuations when a similar up and down curve could be observed. While if using only citations, such curve is hardly seen.

Table~\ref{tab.perf} gives the detailed quality scores of these three clustering solutions based on the pseudo-ground-truth. The last column gives the average Adjusted Mutual Information scores (AMI)~\cite{nami} between this solution and the other four solutions, namely CWTS-C5, UMSI0, ECOOM-BC13 and STS-RG. We see that if using only citations, the resulting clusters agree with the other clustering solutions more than those if no citations are used whose $adjustedF1$ score is also the lowest. It is not surprising as the other clustering solutions rely heavily on the citation information. So, even if the ways of using citations are different, the citation information still brings enough agreement between them. Using all entities to represent articles has the highest $adjustedF1$ score and agrees with the others the most. 

Table~\ref{tab.ami1} gives the AMI scores between these three solutions and the clusters based on the Louvain method. Again clusters based on only citations agree with the Louvain results almost to the same degree as those using all entities do. According to these measures, we decided to use all entities as the final selection of features, and keep these 31 clusters as our final K-Means results, labelled as OCLC-31. The size distribution of these 31 clusters is shown in Fig~\ref{fig.size} (a). 

\begin{table}[t]
\centering
\caption{Quality comparison among different feature selections}
\label{tab.perf}       
\begin{tabular}{rcccccc}\hline
 &\#clusters& r & p & f1 & $adjustedF1$& Average AMI to others\\\hline
 no citations & 24 & 0.53 & 0.17 & 0.26 & 2.16 & 0.44\\
 only citations & 28 & 0.58 & 0.18 & 0.28 & 2.29 &0.47\\
 all entities & 31 & 0.56 & 0.23 & 0.33 & 2.69 & 0.47\\\hline
 oclc\_louvain  &32& 0.61& 0.21  & 0.31 & 2.57 &0.49 \\\hline
\end{tabular}
\end{table}

%\begin{table}[t]
%\centering
%\caption{Quality comparison among different feature selections}
%\label{tab.perf2}       
%\begin{tabular}{rccccc}\hline
%&\#clusters& r & p & f1 & $adjustedF1$\\\hline
%no citations & 24 & 0.48 & 0.23 & 0.31 & 2.60\\
%only citations & 28 & 0.54& 0.25  & 0.34 &2.83\\
%all entities& 31 & 0.52 & 0.32 & 0.39 & 3.20\\\hline
%oclc\_louvain & 32&0.60 & 0.31 & 0.41 & 3.33\\\hline
%\end{tabular}
%\end{table}

\begin{table}[t]
\centering
\caption{Adjusted Mutual Information between solutions \label{tab.ami1}}
\begin{tabular}{rcccc}\\\hline
& no citations & only citations & all entities& louvain\\\hline
no citations & 1.00 & 0.59 & 0.63 & 0.56 \\
only citations & &1.00 & 0.69 & 0.65 \\
all entities & &&1.00 & 0.67 \\
oclc\_louvain&&&& 1.00 \\\hline
\end{tabular}
\end{table}

% \begin{table}[t]
% \centering
% \caption{Performance comparison among different feature selections}
% \label{tab.perf}       
% \begin{tabular}{rcccc}\hline
%  & p & r & f & $adjustedF1$\\\hline
%  oclc\_31 & 0.57 & 0.23 & 0.33 &  2.70 \\
%  oclc\_louvain  & 0.21 & 0.62 & 0.32 & 2.58 \\\hline
% \end{tabular}
% \end{table}

\subsection{Community detection using the Louvain method}
Different from the standard application of the Louvain method, whose input is a citation-based network, we apply the Louvain method on a semantic similarity network where each node is an article and there is an edge between two articles if they are highly similar/related. Based on the experiments with K-Means, we again use all entities to compute the semantic representation of articles. For each article, we calculated the top 40 most similar articles whose similarity values are higher than a certain threshold (in this case 0.6) and consider this article and its top 40 closest peers are connected. Once every article is connected to its peers, a similarity network is formed and then it becomes rather straightforward to apply the Louvain method to detect communities or clusters in this network.

\begin{figure}[t]
\centering 
\begin{tabular}{c}
\includegraphics[width=.7\linewidth]{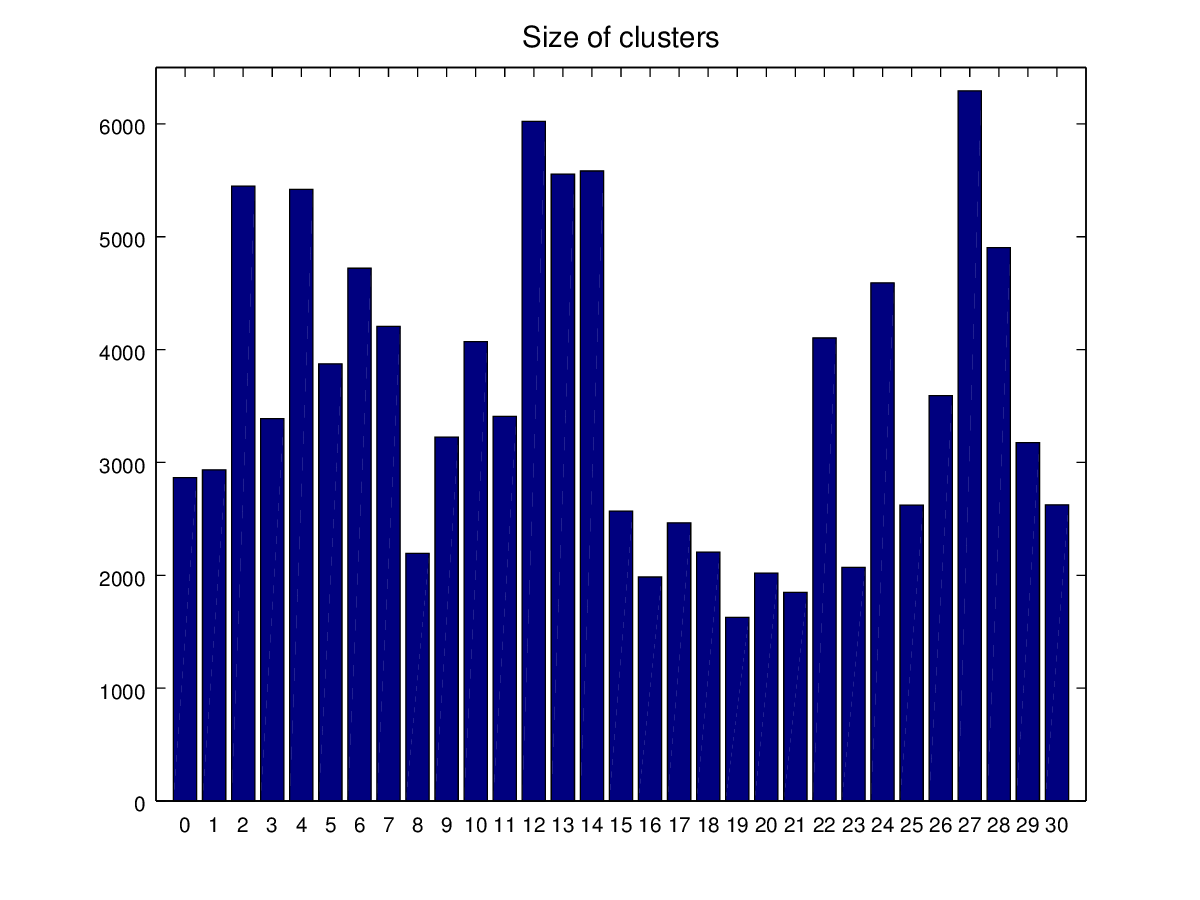}\\
(a) OCLC-31\\
\includegraphics[width=.7\linewidth]{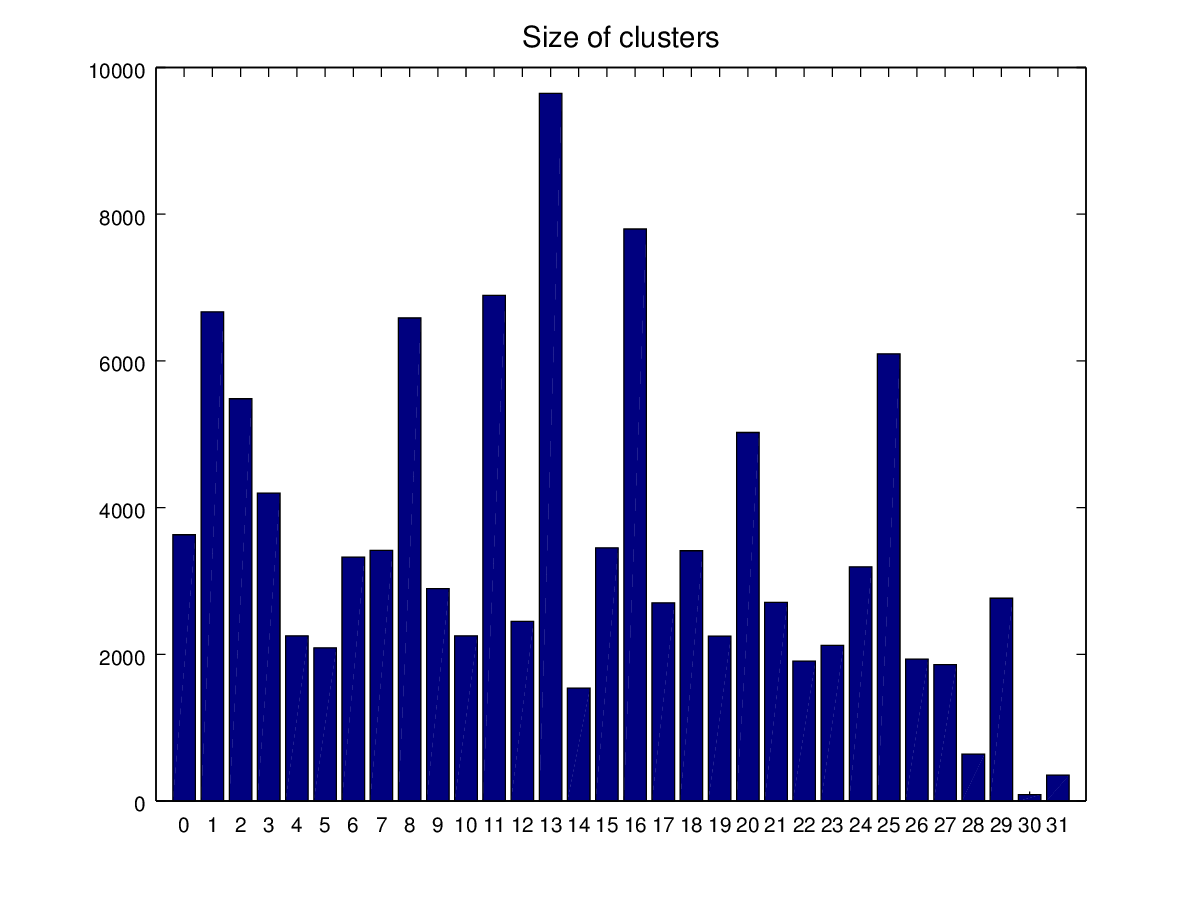}\\
(b) OCLC-Louvain\\
\end{tabular}
\caption{The size distribution of our two clustering solutions.\label{fig.size}}
\end{figure}

We use the python library networkx\footnote{\url{https://networkx.github.io/}} and its community detection module which implements community detection using the louvain method.\footnote{\url{http://perso.crans.org/aynaud/communities/}} This results in 32 best partitions (clusters), labelled as OCLC-Louvain, with the largest partition containing 9646 articles, the smallest 86 articles and in average 3488 articles, see Figure~\ref{fig.size} (b). Its quality against the pseudo-ground-truth is given in Table~\ref{tab.perf}.

The Louvain clusters perform similarly to the K-Means clusters, and actually agrees more with the other clustering solutions. However, the disadvantage of using the Louvain method is that it is not scalable for a bigger dataset as the similarity network is expensive to generate using a distance metric, even if the Louvain algorithm itself is relatively scalable. 

\subsection{Consensus checking}

Now we can use standard consensus measures such as Adjusted Mutual Information (AMI)~\cite{nami} to check how much these two clustering solutions agree with each other. Table~\ref{tab.ami2} gives the consensus score between these two solutions and the other five solutions reported in this special issue.\footnote{The CWTS-C5 and UMSI0 are the clustering solutions generated by two different methods, Infomap and the Smart Local Moving Algorithm (SLMA) respectively, applied on the direct citation network of articles. The two ECOOM clustering solutions are generated by applying the Louvain method to find communities among bibliographic coupled articles where ECOOM-NLP11 also incorporates the keywords information. The STS-RG clusters are generated by first projecting the small \textit{Astro} dataset to the full Scopus database and collecting their cluster assignments after the full Scopus articles are clustered using SLMA on the direct citation network. More detailed account can be found in~\cite{velden2015comparison}. } The last row gives the average AMI between one clustering solution and all the other solutions.   
\begin{table}[h]
\centering
\caption{Consensus checking using Adjusted Mutual Information (AMI)}
\label{tab.ami2}
\begin{tabular}{rrrrrrrr}\\\hline
	&sr & c& u & eb & en & ok & ol \\\hline
STS-RG (sr)& 1.0 & 0.44 & 0.46 & 0.43 & 0.34 & 0.41 & 0.42 \\
CWTS-C5 (c) & & 1.0 & 0.77 & 0.47 & 0.39 & 0.56 & 0.61 \\
UMSI0 (u) & & & 1.0 & 0.47 & 0.38 & 0.51 & 0.55 \\
ECOOM-BC13 (eb)& &&&1.0 & 0.46 & 0.46 & 0.46 \\
ECOOM-NLP11 (en) & &&&&1.0 & 0.41 & 0.39 \\
OCLC-31 (ok) & &&&&&1.0 & 0.67 \\
OCLC-Louvain (ol)&&&&&&&1.0\\\hline
Average AMI & 0.42 & 0.54 & 0.52 & 0.46 & 0.40 & 0.50 & 0.52 \\\hline
\end{tabular}
\end{table}

% \begin{table}[t]
% \centering
% \caption{Consensus checking using Adjusted Mutual Information (AMI)}
% \label{tab.ami}
% \begin{tabular}{rrrrrrrrrr}\\\hline
% 	&sts & cwts & umsi & ecoom\_bc & ecoom\_nlp & no citations & only citations & oclc\_k & oclc\_l \\\hline
% sts& 1.00 & 0.44 & 0.46 & 0.43 & 0.34 & 0.41 & 0.40 & 0.41 & 0.42 \\
% cwts & &1.00 & 0.77 & 0.47 & 0.39 & 0.50 & 0.57 & 0.56 & 0.61 \\
% umsi & & & 1.00 & 0.47 & 0.38 & 0.46 & 0.51 & 0.51 & 0.55 \\
% ecoom\_bc13& &&& 1.00 & 0.46 & 0.44 & 0.47 & 0.46 & 0.46 \\
% ecoom\_nlp11& &&&& 1.00 & 0.41 & 0.41 & 0.41 & 0.39 \\
% no citations &&&&&& 1.00 & 0.59 & 0.63 & 0.56 \\
% only citations &&&&&&& 1.00 & 0.69 & 0.65 \\
% oclc\_k& &&&&&&&1.0 & 0.67 \\
% oclc\_l&&&&&&&&&1.0\\\hline
% \end{tabular}
% \end{table}

These numbers suggest that the data model has more impact on the solution than the algorithm chosen because OCLC-31  and OCLC-Louvain have the second highest value in terms of agreement with each other (the highest agreement is between CWTS-C5 and UMSI0, which also use the same data model). Comparing to STS-RG and ECOOM solutions, our two solutions agree more with CWTS-C5 and UMSI0, which indicates that even with a different data model, the results are still highly comparable. More detailed comparison can be found in~\cite{velden2015comparison}.

\section{Conclusion}
In this paper, we applied two clustering methods to identify clusters in the Astro dataset. Different from the other methods presented in this special issue, we built semantic representation for articles and tried to detect clusters of articles based on their semantic similarity. We gave technical details and the decision path towards our two clustering solutions, one based on K-Means and one based on Louvain community detection method. 

The semantic representation of articles is built on a semantic matrix to which these articles contribute. Each entity (topical terms, subject, author, journal, citation) is represented by its lexical environment extracted and highly reduced from the corpus. We integrated the semantic vectors of all entities involved in one article as the representation of this article. Our experiments show that such integration of the semantics of the individual entities reflects the semantics of articles and the clustering results are competitive with other clustering solutions which are mainly based on citation information. 

We would like to emphasise that the two clustering methods used in this paper are only two options we tried on such semantic representation. K-Means is highly scalable and produces results with high agreement with other solutions. One advantage is that it is applicable when citation data is missing. It could be a first step of clustering to separate articles based on their lexical information, before diving into relevant subsets with more delicate and complex clustering methods.  

\label{sec.conclusion}

\textbf{Acknowledgement}: Part of this work has been funded by the COST Action TD1210 Knowescape. We would like to thank Jochen Gl\"aser and Andrea Scharnhorst for extended comments on earlier versions of the text. We would also like to thank the internal reviewer Michael Heinz as well as the anonymous external referees for their valuable comments and suggestions. 

% BibTeX users please use one of
%\bibliographystyle{spbasic}      % basic style, author-year citations
\bibliographystyle{spmpsci}      % mathematics and physical sciences
\bibliography{astro.bib}   % name your BibTeX data base

\end{document}